\documentstyle[12pt, epsfig]{article}
\newcommand{\be}{\begin{equation}}
\newcommand{\ee}{\end{equation}}
\newcommand{\bi}{\begin{itemize}}
\newcommand{\ei}{\end{itemize}}
\newcommand{\bfi}{\begin{figure}}
\newcommand{\efi}{\end{figure}}
\newcommand{\bc}{\begin{center}}
\newcommand{\ec}{\end{center}}

\newcommand{\bea}{\begin{eqnarray}}
\newcommand{\eea}{\end{eqnarray}}

\newcommand{\xn}{\frac{\theta}{2\pi i}}

\newcommand{\bdm}{\begin{displaymath}}
\newcommand{\edm}{\end{displaymath}}
\newcommand{\bit}{\begin{itemize}}
\newcommand{\eit}{\end{itemize}}
\newcommand{\arb}{\begin{array}}
\newcommand{\are}{\end{array}}
\def\ep {\epsilon}

\def\th {\theta}
\def\kg{\Gamma}

\newfont{\rt}{cmr10}
\newcommand{\ket}[1]{|{#1} \rangle}

\newcommand{\psbildc}[5]
{\par
 \begin{figure}[#1]
 \bc
 \begin{minipage}{0cm} \end{minipage}
 \begin{minipage}{#3}
 \refstepcounter{figure}\label{#2}
 \epsfxsize=#3
 \epsffile{#4}
 \end{minipage}
 \ec 
 \hfill
 \begin{minipage}{0cm} \end{minipage}
 \bc
 \parbox{15cm}{\centerline {\baselineskip12pt{\rt {\bf fig. \ref{#2}:} #5}}}
 \ec
 \end{figure}}

\newcommand{\half}{\frac{1}{2}}
\begin{document}
\newcommand{\nd}[1]{/\hspace{-0.5em} #1}
\def\rar{\rightarrow}

\begin{titlepage}
\begin{flushright}
SWAT/166\\
hep-th/9710017 \\
**** 1997  \\
\end{flushright}
\vspace{.6in}
\begin {centering}
{Topological excitations in  $N=1$ Supersymmetric QFT.}

\vspace{.12in}
 Evangelos Mavrikis$^a$\\
\vspace{.1in}
Department of Physics, University of Wales Swansea \\
Singleton Park, Swansea, SA2 8PP, UK \\
\vspace{1.4in}
{\bf Abstract} \\
The present paper deals  with $N=1$ 2D  supersymmetric integrable quantum field theory. The S-matrix proposed to describe the interactions between  supersymmetric particles is  applied to theories involving topological excitations of zero central charge. Bound states can fit  consistently   within this type of  theories, since the bootstrap can be shown to close. The topological character of the excitations and the similarity with the scattering of particles is fully understood when a kink sector is introduced in the theory.
\end{centering}
{\small }
\vspace{2cm}
\begin{flushleft}
$^a$pyem@swanea.ac.uk
\end{flushleft}

\end{titlepage}
\section{Introduction}

In a resent work \cite{hm},  based on the original work of Schoutens \cite{sch}, the bootstrap equation was investigated in 1+1 integrable quantum field theories involving  supersymmetric   particles. The presence of a kink sector and its implications in such theories  was considered  as well. Particles were considered to fall into representations of $N=1$ supersymmetry with zero central charge, while kinks carry  non  zero central charge. In  addition, kinks  are subject to an adjacency condition, reflecting the  non trivial topology of the non-zero charge  sector.

Schoutens has also discussed a very interesting case concerning excitations of  a mixed character. They are carries of a  SUSY representation with no central charge, but they also obey an adjacency condition. Despite the last crucial difference, the  S-matrix proposed to describe the interactions between  those excitations appears to be very similar with the one for particles. The present work is an attempt to explore theories involving  such excitations and discover the origin of the above formal similarity.

This letter  is organised as follows. In section 2 some aspects of supersymmetric theories involving  particles are reviewed. Then Schoutens's theory for topological excitations is generalized to involve particles of  different mass. Bootstrap is  discussed in those theories as well. In section 3, a kink sector is introduced and the way it fits within  theories of particles is shortly reviewed. The topological excitations are also considered in the precense of such a sector  and their topological character  is then  explained. Finally, in section 4 the problem of diagonalizing the fermion parity is discussed.

\section{Scattering of  supersymmetric excitations of zero central charge}
\subsection {S-Matrix for super-particles}
In 1+1 dimensions the basis of  $N=1$ SUSY irreducible representation with zero central charge  consists of two states  $\{ \ket{\phi (\th)}, \ket{\psi (\th)}\}$ of mass $m$ ($\th$ is the rapidity of the particle). In this basis, the supercharges take the matrix form
\be
{\cal Q}=e^{\th /2}\sqrt{m}\pmatrix {0&\ep \cr \ep^{*} &0 \cr },~~ \bar{\cal Q}=e^{-\th /2}\sqrt{m}\pmatrix {0&\ep ^* \cr \ep  &0 \cr },~~Q_L=\pmatrix{1&0 \cr 0&-1\cr}  
\ee
where $\ep=\exp(i\pi/4)$ and  $Q_L$ is the fermionic parity operartor.
The action of supercharges on two-particle states $\ket{A_1(\th _1)A_2(\th _2)}=\ket{A_1(\th _1)}\otimes \ket{A_2(\th _2)}$ is
\be
\Delta({\cal Q})={\cal Q}\otimes I+Q_L \otimes {\cal Q},~~\Delta(\bar{\cal Q})=\bar{\cal Q}\otimes I+Q_L \otimes \bar{\cal Q} \label{mua}
\ee

Under the assumption of integrability, it is possible to construct a minimal  S-matrix $S_P(\th)$ for a QFT involving superparticles in the above representation. This construction is based on the commutation with supercharges and the  requirements for  unitarity, crossing symmetry and the Yang-Baxter equation \cite{sch}. Bound states can be introduced by multiplying $S_P(\th)$ with a purely bosonic consistent S-matrix $S_B(\th)$ that does  exhibit  poles at particular imaginary values of rapidity difference \cite{cor1}. The additional requirement for closing the bootstrap  implies strong restrictions on  the spectrum of the full theory described by $S_P(\th)\otimes S_B(\th)$.

The masses of the doublets have to be of the form
\be
m_a=m\sin(a\pi/H),~a=1,2,...,n \label{sp}
\ee
where $n$ is the total number of particles. If $H=2n$, the particles are not self-conjugate. For simplicity, the particles will be taken to be self-conjugate. The fusing angles must also obey a specific rule:
\be
u_{ab}^{a+b}=\frac{(a+b)\pi}{H}~(a+b\leq n),~~~~u_{ab}^{|b-a|}=\pi-\frac{|b-a|\pi}{H},~(a+b>n) \label{fc1}
\ee
In this case the elements of $S_P(\th)$ take the form 
\bea
S_{\phi\phi
\rightarrow\phi\phi}^{[ab]}(\th)&=&\left(1+{2\sin({a+b\over 2H}\pi)\cos(
{a-b\over 2H}\pi)\over \sin({\theta\over i})} \right)g^{[ab]}(\theta), \nonumber \\ \nonumber \\
S_{\psi \psi\rightarrow\psi\psi}^{[ab]}
(\theta)&=&\left(-1+{2\sin({a+b\over 2H}\pi)\cos({a-b\over 2H}\pi)
\over\sin({\theta\over i})}\right)g^{[ab]}(\theta) \nonumber \\ \nonumber \\ 
S_{\phi\phi\rightarrow\psi\psi}^{[ab]}(\theta)&=&S_{\psi 
\psi\rightarrow\phi\phi}^{[ab]}(\theta)=
{\sqrt{\sin({a\pi\over H})\sin({b\pi\over H})}\over \cos({\theta\over 2i})}
g^{[ab]}(\theta),\nonumber \\ \label{ap} \\
S_{\phi\psi\rightarrow\phi\psi}^{[ab]}(\theta)&=&
S_{\psi\phi\rightarrow\psi\phi}^{[ab]}(\theta)=
{\sqrt{\sin({a\pi\over H})\sin({b\pi\over H})}\over 
\sin({\theta\over2i})}g^{[ab]}(\theta),\nonumber \\ \nonumber \\
S_{\phi\psi\rightarrow\psi\phi}^{[ab]}(\theta)&=&
\left(1-{2\sin({a-b\over 2H}\pi)\cos(
{a+b\over 2H}\pi)\over \sin({\theta\over i})}
\right)g^{[ab]}(\theta), \nonumber \\ \nonumber \\
S_{\psi\phi\rightarrow\phi\psi}^{[ab]}(\theta)&=&
\left(1+{2\sin({a-b\over 2H}\pi)\cos(
{a+b\over 2H}\pi)\over \sin({\theta\over i})}
\right)g^{[ab]}(\theta).\nonumber
\eea
The functions $g^{ab}(\th)$ are fixed by unitarity and crossing symmetry:
\bea 
g^{[ab]}(\th)&=&R^{[ab]}(\th)R^{[ab]}(i\pi-\th),\nonumber \\ \nonumber \\
R^{[ab]}(\th)&=&{1\over \kg(\xn)\kg(\xn+\half)}  \prod_{k=1}^{\infty}
{\kg(\xn +{a+b\over2H}+k-1)  \kg(\xn - {a+b\over2H} +k)\over \kg(\xn
+{a+b\over2H} +k-\half) \kg(\xn -{a+b\over2H}+k+ \half)} \nonumber \\ \nonumber \\  
&\qquad&\times {\kg(\xn + {a-b\over2H} +k- \half ) \kg( \xn -{a-b\over2H}+k
-\half)\over  \kg( \xn +{a-b\over2H} +k) \kg( \xn -{a-b\over2H} +k)}.
\eea

The fusion has the form
\bea
\ket{\phi^a (\th+i\bar{u}_{a\bar{c}}^{\bar{b}})\phi^b (\th+i\bar{u}_{b\bar{c}}^{\bar{a}})}&=&(f_{\phi \phi})_{ab}^c\ket{\phi ^c(\th)}\nonumber \\
\ket{\psi^a (\th+i\bar{u}_{a\bar{c}}^{\bar{b}})\psi^b (\th+i\bar{u}_{b\bar{c}}^{\bar{a}})}&=&(f_{\psi \psi})_{ab}^c\ket{\phi ^c(\th)}\nonumber \\
\ket{\phi^a (\th+i\bar{u}_{a\bar{c}}^{\bar{b}})\psi^b (\th+i\bar{u}_{b\bar{c}}^{\bar{a}})}&=&(f_{\phi \psi})_{ab}^c\ket{\psi ^c(\th)}\label{bs} \\
\ket{\psi^a (\th+i\bar{u}_{a\bar{c}}^{\bar{b}})\phi^b (\th+i\bar{u}_{b\bar{c}}^{\bar{a}})}&=&(f_{\psi \phi})_{ab}^c\ket{\psi ^c(\th)}\nonumber
\eea
where
\bea 
(f_{\phi \phi})_{ab}^c=\sqrt{S^{ab}_{\phi \phi \rar \phi \phi}(iu_{ab}^c)},&&~~(f_{\psi \psi})_{ab}^c=\sqrt{S^{ab}_{\psi \psi \rar \psi \psi}(iu_{ab}^c)} \nonumber \\
(f_{\phi \psi})_{ab}^c=\sqrt{S^{ab}_{\phi \psi \rar \psi \phi}(iu_{ab}^c)},&&~~(f_{\psi \phi})_{ab}^c=\sqrt{S^{ab}_{\psi \phi \rar \phi \psi}(iu_{ab}^c)}\label{fsm}
\eea

Notice at this stage that the same particle $c$ can be represented equally well by {\em any} one of the $b,c$ breathers
\bea
\ket{\phi ^c(\th)}&=&\frac{1}{(f_{\phi \phi})_{ab}^c}\ket{\phi^a (\th+i\bar{u}_{a\bar{c}}^{\bar{b}})\phi^b (\th+i\bar{u}_{b\bar{c}}^{\bar{a}})}=\frac{1}{(f_{\psi \psi})_{ab}^c}\ket{\psi^a (\th+i\bar{u}_{a\bar{c}}^{\bar{b}})\psi^b (\th+i\bar{u}_{b\bar{c}}^{\bar{a}})}\nonumber \\ \label{ibs}\\
\ket{\psi ^c(\th)}&=&\frac{1}{(f_{\phi \psi})_{ab}^c}\ket{\phi^a (\th+i\bar{u}_{a\bar{c}}^{\bar{b}})\psi^b (\th+i\bar{u}_{b\bar{c}}^{\bar{a}})}=\frac{1}{(f_{\psi \phi})_{ab}^c}\ket{\psi^a (\th+i\bar{u}_{a\bar{c}}^{\bar{b}})\phi^b (\th+i\bar{u}_{b\bar{c}}^{\bar{a}})}\nonumber
\eea
 
Consistency of the bootstrap means that  any one of the above expressions  leads to the correct amplitute (\ref{ap}) when it replaces the corresponding particle in a scattering process.

\subsection {S-matrix for topological excitations}

The S-matrix  constructed  by Schoutens involves a set of  four such excitations, all of equal mass $m$, interpolating  between two vacua $1,2$. They will be  denoted as $$\{ \ket{B_{00}(\th)},\ket{B_{10}(\th)}, \ket{B_{11}(\th)},\ket{B_{01}(\th)}   \}$$.
\psbildc{h}{fig1}{8cm}{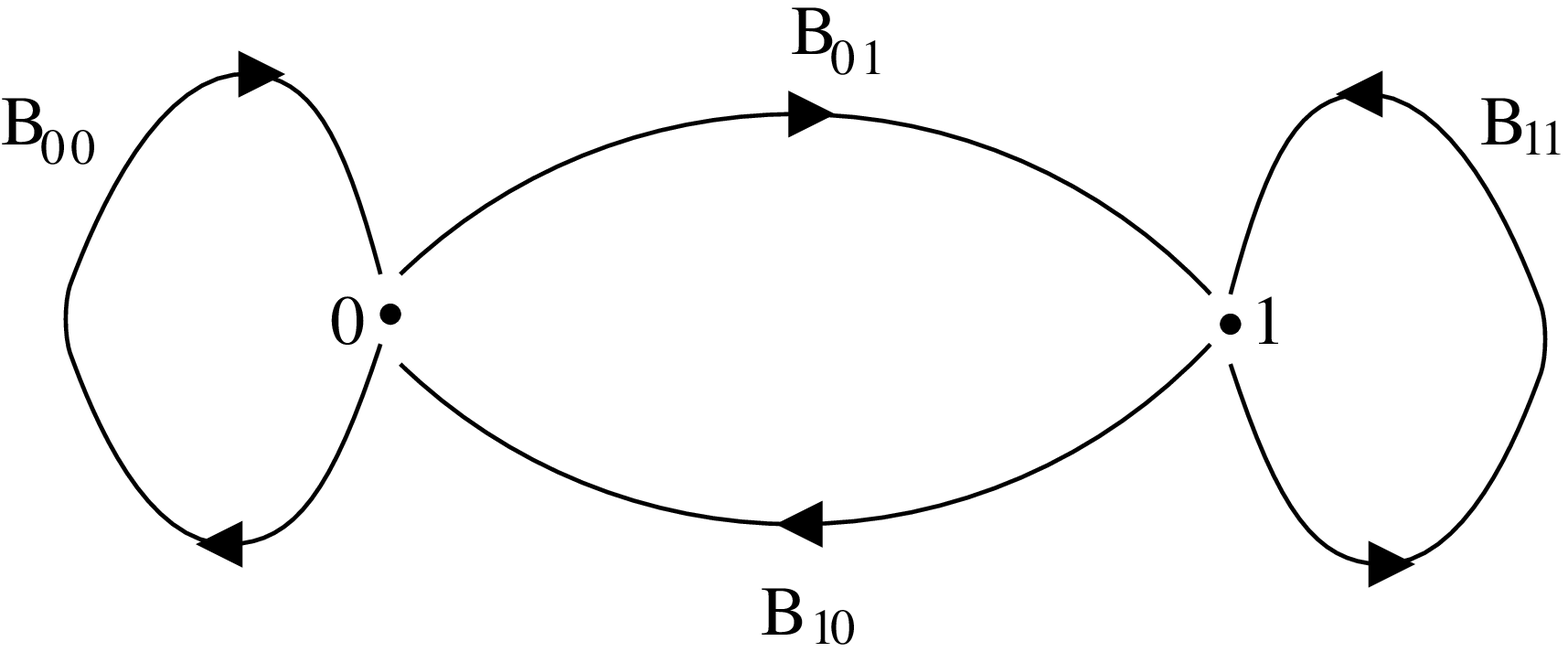}{Topology of the four excitations}

 Obviously they form a reducible representation, consisting of two irreducible ones joined by the fermion parity:
\bea
Q_L&=&\pmatrix{0&0&1&0 \cr 0&0&0&-1 \cr 
1&0&0&0\cr 0&-1&0&0} ,~~
{\cal Q}=e^{\th /2}\sqrt{m}\pmatrix{0&e^{-i\pi /4}&0&0 \cr e^{i\pi /4}&0&0&0 \cr 0&0&0&e^{-i\pi /4} \cr 0&0&e^{i\pi /4}&0}  \nonumber \\
\nonumber \\ \\
\bar{\cal Q}&=&e^{-\th /2}\sqrt{m}\pmatrix{0&e^{i\pi /4}&0&0 \cr e^{-i\pi /4}&0&0&0 \cr 0&0&0&e^{i\pi /4} \cr 0&0&e^{-i\pi /4}&0}  \nonumber 
\eea
There are only eight  admissable  two-particle states. An S-matrix element for the process $B_{\alpha \beta}(\th _1)+B_{\beta \gamma}(\th _2) \rar B_{\alpha \delta}(\th _2)+B_{\delta  \gamma}(\th _1)$ will be denoted by
    $$S\left(\left.\matrix{\alpha&\delta\cr \beta &\gamma}
\right\vert\th _1 -\th _2 \right)$$ 
 The minimal expressions for the  supersymmetric  amplitudes, obeying unitarity, crossing symmetry and the Yang-Baxter equation, are:
\bea
S\left(\left.\matrix{0&0\cr 0 &0\cr}
\right\vert\th\right)&=&S\left(\left.\matrix{1&1\cr 1 &1\cr}
\right\vert\th\right)=\left(1+\frac{2}{\sin(\th /i)}\right)g(\th) \nonumber\\
\nonumber \\
S\left(\left.\matrix{0&1\cr 1 &0\cr}
\right\vert\th\right)&=&S\left(\left.\matrix{1&0\cr 0 &1\cr}
\right\vert\th\right)=\left(-1+\frac{2}{\sin(\th /i)}\right)g(\th) \nonumber \\
\nonumber \\
S\left(\left.\matrix{0&0\cr 1 &0\cr}
\right\vert\th\right)&=&S\left(\left.\matrix{1&1\cr 0 &1\cr}
\right\vert\th\right)=S\left(\left.\matrix{1&0\cr 1 &1\cr}
\right\vert\th\right)=S\left(\left.\matrix{0&1\cr 0 &0\cr}
\right\vert\th\right)= -\frac{1}{\cos(\th /2i)}g(\th) \nonumber \\ \\
S\left(\left.\matrix{0&1\cr 1 &1\cr}
\right\vert\th\right)&=&S\left(\left.\matrix{0&0\cr 0 &1\cr}
\right\vert\th\right)=S\left(\left.\matrix{1&1\cr 1 &0\cr}
\right\vert\th\right)=S\left(\left.\matrix{1&0\cr 0 &0\cr}
\right\vert\th\right)= \frac{1}{\sin(\th /2i)}g(\th) \nonumber \\
\nonumber\\
S\left(\left.\matrix{0&0\cr 1 &1\cr}
\right\vert\th\right)&=&S\left(\left.\matrix{1&1\cr 0 &0\cr}
\right\vert\th\right)=S\left(\left.\matrix{0&1\cr 0 &1\cr}
\right\vert\th\right)=S\left(\left.\matrix{1&0\cr 1 &0\cr}
\right\vert\th\right)= g(\th) \nonumber 
\eea

where 
\be
g(\th)=R(\th)R(i\pi-\th),~
R(\th)=\frac{1}{\kg(\frac{\th}{2\pi i})\kg(\frac{\th}{2\pi i}+\half)}\prod_{k=1}^{\infty}\left( \frac{\kg(\frac{\th}{2 \pi i}-\half +k)}{\kg(\frac{\th}{2 \pi i} +k)} \right)^4 \nonumber
\ee
Crossing symmetry reads 
\be
\ket{\overline{B_{00}}(\th)}=\ket{B_{00}(\th)},~\ket{\overline{B_{01}}(\th)}=\ket{B_{10}(\th)},~\ket{\overline{B_{11}}(\th)}=\ket{B_{11}(\th)},~\ket{\overline{B_{10}}(\th)}=\ket{B_{01}(\th)} \nonumber
\ee
It can be easily seen that these amplitudes coincide  with the ones in (\ref{ap}) for $a=b=H/2$. At this stage one wonders if, in the spirit of  this  similarity, the spectrum can be enlarged to include  supermultimpletes of different mass. The S-matrix elements for such a theory will be the same as in (\ref{ap}). The aim is to use these amplitudes  in such a way that unitarity, crossing symmetry and the  Young-Baxter are not spoiled by the presence of the adjacency condition.

Consider then  $n$ four-dimensional multipletes $\{ \ket{B_{00}^a(\th)},\ket{B_{10}^a(\th)}, \ket{B_{11}^a(\th)},\ket{B_{01}^a(\th)}   \}$  of mass $m_a,~a=1,2,...,n$. Keeping in advance an eye on the bootstrap , it is a natural choice for the spectrum to be of the form (\ref{sp}). It is not  a hard  guess to chose:
\bea
S^{[ab]}\left(\left.\matrix{0&0\cr 0 &0\cr}
\right\vert\th\right)&=&S^{[ab]}\left(\left.\matrix{1&1\cr 1 &1\cr}
\right\vert\th\right)=S^{[ab]}_{\phi \phi \rar \phi \phi}(\th) \nonumber\\
\nonumber \\
S^{[ab]}\left(\left.\matrix{0&1\cr 1 &0\cr}
\right\vert\th\right)&=&S^{[ab]}\left(\left.\matrix{1&0\cr 0 &1\cr}
\right\vert\th\right)=S^{[ab]}(\th)_{\psi \psi \rar \psi \psi} \nonumber \\
\nonumber \\
S^{[ab]}\left(\left.\matrix{0&0\cr 1&0\cr}
\right\vert\th\right)&=&S^{[ab]}\left(\left.\matrix{1&1\cr 0 &1\cr}
\right\vert\th\right)=S^{[ab]}\left(\left.\matrix{1&0\cr 1 &1\cr}
\right\vert\th\right)=S^{[ab]}\left(\left.\matrix{0&1\cr 0 &0\cr}
\right\vert\th\right)=\left\{ \matrix{~ S^{[ab]}(\th)_{\phi \phi \rar \psi \psi}\cr S^{[ab]}(\th)_{\psi \psi \rar \phi \phi}} \right. \nonumber \\\label{am2} \\
S^{[ab]}\left(\left.\matrix{0&1\cr 1 &1\cr}
\right\vert\th\right)&=&S^{[ab]}\left(\left.\matrix{0&0\cr 0 &1\cr}
\right\vert\th\right)=S^{[ab]}\left(\left.\matrix{1&1\cr 1 &0\cr}
\right\vert\th\right)=S^{[ab]}\left(\left.\matrix{1&0\cr 0 &0\cr}
\right\vert\th\right)= \left\{ \matrix{S^{[ab]}(\th)_{\phi \psi \rar \phi \psi} \cr S^{[ab]}(\th)_{\psi \phi \rar \psi \phi}} \right.  \nonumber \\
\nonumber\\
S^{[ab]}\left(\left.\matrix{0&0\cr 1 &1\cr}
\right\vert\th\right)&=&S^{[ab]} \left(\left.\matrix{1&1\cr 0 &0\cr}
\right\vert\th\right)=S^{[ab]}(\th)_{\phi \psi \rar \psi \phi} \nonumber \\ \nonumber \\
S^{[ab]}\left(\left.\matrix{0&1\cr 0 &1\cr}
\right\vert\th\right)&=&S^{[ab]}\left(\left.\matrix{1&0\cr 1 &0\cr}
\right\vert\th\right)= S^{[ab]}(\th)_{\psi \psi \rar \phi \psi} \nonumber 
\eea

 Unitarity and crossing symmetry can be clearly seen, while  Yang-Baxter needs some more, but really  straight forward calculation.

Naturally, the next step is to explore the possibility of introducing bound states by attaching a purely bosonic piece to the minimal supersymmetric S-matrix. The adjacency  implies the following form for the fusion:
$$\ket{B^a_{\alpha \beta}(\th+iu_{a\bar{c}}^{\bar{b}})B^b_{\beta \gamma}(\th-iu_{b\bar{c}}^{\bar{a}}) }=(f_{\alpha \beta \gamma}^{\alpha \gamma})_{ab}^c \ket{B_{\alpha \gamma}^c (\th)} $$ where the greek indices label one of the two vacuua of the theory.

Notice that there are  again  two breathers corresponding to the same particle state:
\be
\ket{B_{\alpha \gamma}^c (\th)}=\frac{1}{(f_{\alpha \mu \gamma}^{\alpha \gamma})_{ab}^c} \ket{B^a_{\alpha \mu}(\th+iu_{a\bar{c}}^{\bar{b}})B^b_{\mu \gamma}(\th-iu_{b\bar{c}}^{\bar{a}}) },~~\mu=0,1 \label{br1}
\ee
The fusion coupling are determined by the S-matrix elements:
\be
(f_{\alpha \mu \gamma}^{\alpha \gamma})_{ab}^c=\sqrt{S^{[ab]}\left(\left.\matrix{\alpha &\mu\cr \mu  &\gamma\cr}
\right\vert iu_{ab}^c\right)} \label{fcsm}
\ee
Working out the above relation one finds :
\bea
(f_{000}^{00})_{ab}^c=(f_{111}^{11})_{ab}^c=(f_{\phi \phi})_{ab}^c~~,(f_{010}^{00})_{ab}^c=(f_{101}^{11})_{ab}^c=(f_{\psi \psi})_{ab}^c \nonumber \\ \\ (f_{001}^{01})_{ab}^c=(f_{110}^{10})_{ab}^c=(f_{\phi \psi})_{ab}^c,~ ~(f_{011}^{01})_{ab}^c=(f_{100}^{10})_{ab}^c=(f_{\psi \phi})_{ab}^c  \nonumber
\eea

The bootsrap now reads
\bea
S^{[dc]}\left(\left.\matrix{\alpha &\delta \cr \beta  &\gamma\cr}
\right\vert \th _d -\th _c\right)=\sum_{\nu=0,1}\frac{(f_{\alpha \nu \delta}^{\alpha \delta})_{ab}^c}{(f_{\beta \mu \gamma}^{\beta \gamma})_{ab}^c}~S^{[da]}\left(\left.\matrix{\alpha &\nu \cr \beta  &\mu\cr} \right\vert \th _d -\th _a\right)S^{[db]}\left(\left.\matrix{\nu &\delta \cr \mu  &\gamma\cr} \right\vert \th _d -\th _b\right)\nonumber \\
\mbox{ie.}~~~~~~~~~~~~~~~~~~~~~~~~~~~~~~~~~~~~~~~~~~~~~~~~~~~~  \label{boot1} \\
S^{[dc]}\left(\left.\matrix{\alpha &\delta \cr \beta  &\gamma\cr}
\right\vert \th \right)=\sum_{\nu=0,1}\frac{(f_{\alpha \nu \delta}^{\alpha \delta})_{ab}^c}{(f_{\beta \mu \gamma}^{\beta \gamma})_{ab}^c}S^{[da]}\left(\left.\matrix{\alpha &\nu \cr \beta  &\mu\cr} \right\vert \th-i\bar{u}_{ac}^{b}\right)S^{[db]}\left(\left.\matrix{\nu &\delta \cr \mu  &\gamma\cr} \right\vert \th+i\bar{u}_{bc}^a \right)\nonumber 
\eea
\psbildc{h}{fig2}{12cm}{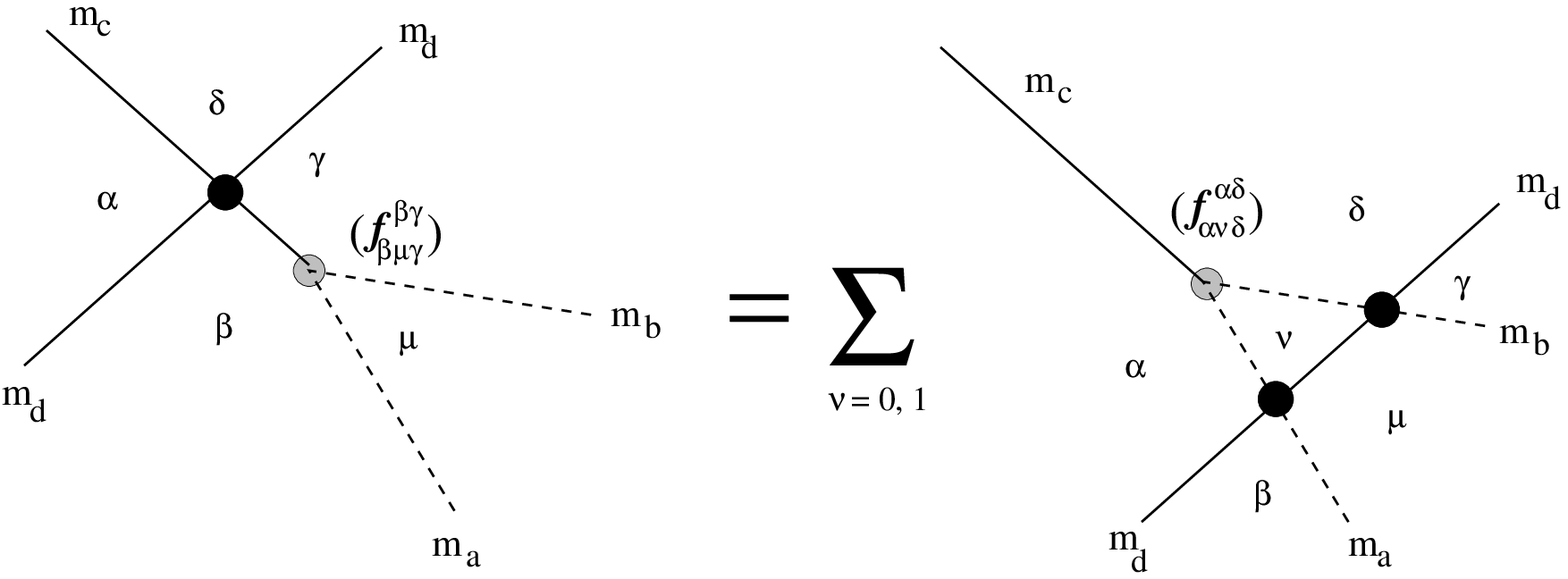}{Bootstrap equation for topological excitations}
Since the S-matrix elements involved are the same as the ones  for particles, the bootstrap can be shown to close in the same fashion as in\cite{hm}. Hence we reach  the conclusion that one can successfully build supersymetric theories involving  excitations of zero central charge and non trivial topology. The adjacency condition is indeed compatible with the S-matrix elements (\ref{am2}). Further  more, bound states can be   introduced  in a successful way that brings  the adjacency condition in full consistency with the  bootstrap.

\section{Introducing the soliton sector}

It is possible that the bosonic spectrum contains particles (or solitons) whose mass and fusing rules  do  not obey  the crucial conditions (\ref{sp}), (\ref{fc1}) respectively. A typical case is the Sine-Gordon theory, where the soliton and antisoliton are of equal mass $m$, while the particles (breathers) have mass
\be
m_a=m\cos(\xi _a /2),~a=1,...,n~~(\xi _a=\pi-2a\pi /H)
\ee

The soliton can fuse with the  antisoliton to any one of the  particles, while soliton (antisoliton) fuse with any  particle back to itself. The fusing angles are
\be
u_{s\bar{s}}^a=u_{\bar{s}s}^a=\pi -\frac{2a\pi}{H},~~~u_{sa}^s=u_{as}^s=u_{\bar{s}a}^{\bar{s}}=u_{a\bar{s}}^{\bar{s}}=\frac{\pi}{2}+\frac{a\pi}{H}
\ee

Clearly, such particles or solitons can not be consistently extended to two-dimensional multipletes of zero central charge. But it is possible to maintain their position in the spectrum of the supersymmetric theory if they are extended to supersymmetric kinks with non zero central charge \cite{hm}, \cite{witten2}. 

The soliton -antisoliton pair is extended to a set of  four kinks interpolating between three vacua , labeled by $\{0,\frac{1}{2},1\}$. In the basis
$\{ \ket{K_{0\half}},\ket{K_{1\half}}, \ket{K_{\half 0}},\ket{K_{\half1}} \}$ the action of the supercharges is realised \cite{ahn1}, \cite{sch}:
\be
{\cal Q}=e^{\th /2}\sqrt{m}\pmatrix{0&i&0&0 \cr -i&0&0&0 \cr 0&0&1&0 \cr 0&0&0&-1},~\bar{\cal Q}=e^{-\th /2}\sqrt{m}\pmatrix{0&i&0&0 \cr -i&0&0&0 \cr 0&0&-1&0 \cr 0&0&0&1},~Q_L=\pmatrix{0&1&0&0 \cr 1&0&0&0 \cr 
0&0&0&1\cr 0&0&1&0} \nonumber
\ee

The kinks  $K_{0\half}$ and $K_{1\half}$ have $T=1$,while their anti-kinks $K_{\half 0}$ and $K_{\half 1}$ have $T=-1$.

\psbildc{h}{kink}{6cm}{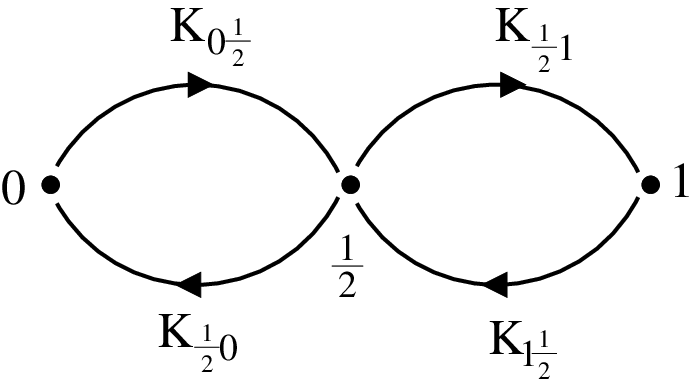}{The four kinks interpolating between the three vacuua}
The non-zero S-matrix  elements are \cite{ahn2}
\bea
S\left(\left.\matrix{0&\half\cr\half &0\cr}
\right\vert\th\right) &=& S\left(\left.
\matrix{1&\half\cr\half &1\cr}\right\vert\th\right)=K(\th)2^{(i\pi-\th)/2\pi
i}\cos\left({\th\over4i}-{\pi\over4}\right)\nonumber \\ \nonumber \\
S\left(\left.\matrix{\half&0\cr0 &\half\cr}\right\vert\th\right)&=&
S\left(\left.
\matrix{\half&1\cr 1&\half\cr}\right\vert\th\right)=K(\th)2^{\th/2\pi
i}\cos\left({\th\over4i}\right)\nonumber \\ \nonumber \\
 S\left(\left.\matrix{0&\half\cr\half &1\cr}\right\vert\th\right)&=&
S\left(\left.
\matrix{1&\half\cr \half&0\cr}\right\vert\th\right)=K(\th)2^{(i\pi-\th)/2\pi
i}\cos\left({\th\over4i}+{\pi\over4}\right)\\ \nonumber\\
S\left(\left.\matrix{\half&1\cr0 &\half\cr}\right\vert\th\right)&=&
S\left(\left.
\matrix{\half&0\cr 1&\half\cr}\right\vert\th\right)=K(\th)2^{\th/2\pi
i}\cos\left({\th\over4i}-{\pi\over2}\right) \nonumber
\eea
The  scalar function $K(\th)$ is determined by crossing symmetry and
unitarity. The minimal solution for $K(\th)$ is
$$
K(\th)={1\over\sqrt\pi}
\prod_{k=1}^{\infty}{ \kg(k-\half +\th /2\pi i)
\kg (k-\th /2\pi i)\over\kg (k+\th /2\pi i)\kg(k+\half -\th /2\pi i)}.
$$

There are  six kink-antikink states which can form  bound states  at  appropriate  (imaginary) values of rapidity. For rapidity difference $\Delta \th=iu_{s\bar{s}}^a$ these  bound states are  related to the particles $\{ \phi^a (\th),\psi^a (\th) \}$:
\be
\ket{K_{\alpha \beta}(\th+i\xi_a/2)K_{\beta \gamma}(\th-i\xi_a/2)}=(f_{\alpha \beta \gamma}^{\phi})_{ss}^a\ket{\phi ^a (\th)}+(f_{\alpha \beta \gamma}^{\psi})_{ss}^a\ket{\psi ^a (\th)}
\ee
The non-zero coupling constants are
\bea
f_{0\half0}^\phi=f_{1\half1}^\phi=2^{(\pi-2\xi)/4\pi}
f_{\half0\half}^\phi=2^{(\pi-2\xi)/4\pi}f_{\half1\half}^\phi=
\sqrt{K(i\xi)2^{(\pi-\xi)/2\pi} 
\cos\left({\xi-\pi\over4}\right)} \nonumber \\
f_{1\half0}^\psi=-f_{0\half1}^\psi=2^{(\pi-2\xi)/4\pi}i
f_{\half0\half}^\psi=-2^{(\pi-2\xi)/4\pi}if_{\half1\half}^\psi=
\sqrt{K(i\xi)2^{(\pi-\xi)/2\pi} 
\cos\left({\xi+\pi\over4}\right)}. \nonumber
\eea

The picture of the particles  $\{ \ket{\phi ^a (\th)},\ket{\psi ^a (\th)} \} $ in terms of the breathers  deserves of special attention. The breathers are still subject to the adjacency condition, while the particles are not. Hence, one needs {\em a set} of  breathers to represent a particle. In the Hilbert space formalism this can be achieved through a summation:
\bea
\ket{\phi ^a(\th)}=\sum_{\alpha, \beta, \gamma}\frac{1}{(f_{\alpha \beta \gamma}^{\phi})_{ss}^a}\ket{K_{\alpha \beta}(\th+i\xi_a/2)K_{\beta \gamma}(\th-i\xi_a/2)} \nonumber \\
\ket{\psi ^a(\th)}=\sum_{\alpha, \beta, \gamma}\frac{1}{(f_{\alpha \beta \gamma}^{\psi})_{ss}^a}\ket{K_{\alpha \beta}(\th+i\xi_a/2)K_{\beta \gamma}(\th-i\xi_a/2)}\label{br0}
\eea

where the parameters $\alpha, \beta, \gamma $ can take the values $0,\half,1$ in a way that respects the topology of the kinks. In fact each one of the particles needs four breathers to be represented. It was first shown in \cite{ahn2} that the breather scattering leads to the correct  amplitudes (\ref{ap}) for the particles, ie the bootstrap closes.

In the case of the zero charge topological excitations, the states  $\{ \ket{B_{\alpha \gamma}^a(\th)},\alpha, \gamma =0,1   \} $  do obey an adjacency condition. It is now obvious that when the solitons are included in the spectrum, each one of the above states can be identified (up to a fusion constant) with a  breather: 
\be
\ket{K_{\alpha \beta}(\th+i\xi_a/2)K_{\beta  \gamma}(\th-i\xi_a/2)}= (f_{\alpha \beta \gamma}^{\phi})_{ss}^a\ket{B_{\alpha \gamma}^a(\th)}\nonumber 
\ee

where $\alpha$, $\gamma$ can only take the values $0,1$.

So, in the presence of a soliton sector, the   the mixed character of the quantum excitations $\ket{B_{\alpha \beta}^a(\th)}$ can be cery well understood in terms of kink-antikink breathers. The fact that particles are represented by a set of the same  breathers also explains the formal similarity of the amplitudes. 

\section{Adjacency condition and fermion parity} 

In the precence of a  non trivial topology, states carry topological indices  related to  the pair of vacua that  the excitation interpolates between. It is this pair  that actually changes under the action of supercharges, hence the fermionic degree is incoded within it. Notice also that the adjacency condition  forces fermion parity to a non diagonal form. When particles are considered, the states are represented by a set of topological breathers, such as adjacency condition is terminated: 
\be
\ket{\phi ^a(\th)}=\frac{1}{\sqrt{2}}(\ket{B_{00}^a(\th)}+\ket{B_{11}^a(\th)}) \nonumber,~~\ket{\psi ^a(\th)}=\frac{1}{\sqrt{2}}(\ket{B_{01}^a(\th)}+\ket{B_{10}^a(\th)})
\ee

In this case, the fermionic degree is released from its topological character and fermion parity is diagonalized. It would be very interesting if one could apply the same idea in the case of solitons. Remember that semiclassical analysis suggests such a picture for supersolitons \cite{kk}, since they appear to be fermion number eigenstates with fractional eigenvalues . In the spirit of diagonalising the fermion parity, the following summation can be attempted for the states of the kink states: 
\bea
\ket{u(\th)}=\frac{1}{\sqrt{2}}(\ket{K_{0\half}(\th)}+\ket{K_{1\half}(\th)}),~~&&\ket{\bar{u}(\th)}=\frac{1}{\sqrt{2}}(\ket{K_{\half 0}(\th)}+\ket{K_{\half 1}(\th)})\nonumber \\ \label{ud} \\
\ket{d(\th)}=\frac{1}{\sqrt{2}}(\ket{K_{0\half}(\th)}-\ket{K_{1\half}(\th)}),~~&&\ket{\bar{d}(\th)}=\frac{1}{\sqrt{2}}(\ket{K_{\half 0}(\th)}-\ket{K_{\half 1}(\th)})\nonumber
\eea

 The adjacency condition has not completely  dissappeared, but it now defines a different topology for the states. The are only two vacua, $A$ (coming from $0,1$) and $B$ ($\half $). States of the same charge interpolate now between the same vacua and transform one to each other under SUSY:
\bea
{\cal Q}\ket{u(\th)}=-ie^{\th /2}\ket{d(\th}&~~~&{\cal Q}\ket{\bar{u}(\th)}=e^{\th /2}\ket{\bar{d}(\th)} \\
{\cal Q}\ket{d(\th)}=+ie^{\th /2}\ket{u(\th}&~~~&{\cal Q}\ket{\bar{d}(\th)}=e^{\th /2}\ket{\bar{u}(\th)}
\eea

 So, in this basis we have achieved a local action for the supercharges and released the fermionic number from the topology. 

\psbildc{h}{fig4}{4cm}{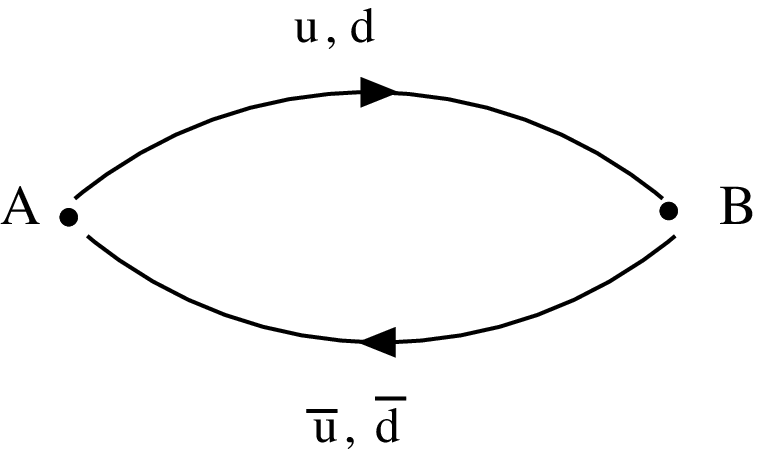}{Topology of the kinks when diagonalise fermion parity}
  
The problem  is  that the above change of basis  is not compatible with crossing symmetry \cite {sch}. I  have attempted to reconstruct a supersymmetric  soliton S-matrix for the states of type (\ref{ud}). Crossing symmetry forces one  to extend the algebra to $N=2$ SUSY in order to gain consistent  braiding for the action on multi-soliton states. Of course there is no problem when $T=0$ (section 2). This  generates the suspicion that  the pathology of the  basis (\ref{ud}) is  somehow related with the  reducibility of the corresponding $N=1$ representation.

Nevertheless, it seems after all that one is oblidged to work with the initial, ''non-diagonal'' basis. The  next step is to gain a realization for the fermion number within this basis. Perhaps the mechanism for  representing a soliton by a set of two kinks is not wrong. It could be simply that the summation rule is not the right way of  applying  the idea. The problem will be investigated extensiveli in future work.

I wish to thank Nadim Mahassen for helping me with my English and C. Ahn for his suggestions.


\begin{thebibliography}{99}

\bibitem{hm}{T. J. Hollowood, E. Mavrikis, Nuclear  Physics B484  (1997), 631} hep-th/9606116
\bibitem{sch}{K.  Schoutens, Nucl. Phys.B 344  (1990) 665}
\bibitem{cor1}{H. W. Braden, E. Corrigan, P.E. Dorey and R. Sasaki, Nucl. Phys. B388, 689 (1990)}
\bibitem{witten2}{E. Witten and D. Olive, Phys. Lett. B78 (1978), 97}
\bibitem{ahn1}{C. Ahn, D. Bernard and A. LeClair, Nucl. Phys. B346 (1990) 409}
\bibitem{ahn2}{C. Ahn, Nucl. Phys. B354 (1991), 57}
\bibitem{lg}{P. Fendley, K. Intriligator, Nucl. Phys.B 373 (1992) 533}
\bibitem{va1}{P. Fendley, S.D. Mathur, C. Vafa and N. P. Warner, Phys. Lett. B243 (1990), 257}
\bibitem{kk}{A. J. Niemi, G. W. Semenoff, Phys. Reports 135 (1986), 99}
\end{thebibliography}
\end{document}